\documentclass[letterpaper]{article} 
\usepackage{aaai25}  
\usepackage{times}  
\usepackage{helvet}  
\usepackage{courier}  
\usepackage[hyphens]{url}  
\usepackage{graphicx} 
\def\UrlFont{\rm}  
\usepackage{natbib}  
\usepackage{caption} 
\usepackage{algorithm}

\usepackage{algpseudocode}
\usepackage{multirow}
\usepackage{lscape}
\usepackage{tabularx}
\usepackage{subcaption}
\usepackage{comment}
\usepackage{amsmath}
\usepackage{amssymb}
\usepackage{lipsum}
\usepackage{multicol}

\usepackage{newfloat}
\usepackage{listings}
\DeclareCaptionStyle{ruled}{labelfont=normalfont,labelsep=colon,strut=off} 
\floatstyle{ruled}
\newfloat{listing}{tb}{lst}{}
\floatname{listing}{Listing}
\title{Federated Learning for Coronary Artery Plaque Detection in Atherosclerosis Using IVUS Imaging: A Multi-Hospital Collaboration}
\author{
    Chiu-Han Hsiao\textsuperscript{\rm 1}, 
    Kai Chen\textsuperscript{\rm 1}, 
    Tsung-Yu Peng\textsuperscript{\rm 2},
    Wei-Chieh Huang\textsuperscript{\rm 3},
}
\affiliations{
    \textsuperscript{\rm 1}Research Center for Information Technology Innovation, Academia Sinica, Taipei City, Taiwan.\\

    \textsuperscript{\rm 2}Department of Information Management, National Taiwan University, Taipei City, Taiwan.\\ 
    \textsuperscript{\rm 3}Division of Cardiology, Department of Internal Medicine, Taipei Veterans General Hospital, School of Medicine, National Yang Ming Chiao Tung University, and Department of Biomedical Engineering, National Taiwan University, Taipei, Taiwan.\\ 

    chiuhanhsiao@citi.sinica.edu.tw, kaichen083@citi.sinica.edu.tw, h1689y@gmail.com, hwcc0314@gmail.com
}

\usepackage{bibentry}

\begin{document}

\maketitle

\begin{abstract}
The traditional interpretation of Intravascular Ultrasound (IVUS) images during Percutaneous Coronary Intervention (PCI) is a time-consuming process that often lacks consistency due to its reliance on the expertise and subjective judgment of physicians. Additionally, the collection and integration of IVUS data across multiple hospital systems are significantly constrained by regulatory restrictions and privacy concerns, posing challenges to collaborative analysis and model development. To establish a consistent and efficient method for IVUS image segmentation and enhance plaque detection during PCI procedures, a parallel 2D U-Net model with a multi-stage segmentation architecture has been developed. The model leverages a federated learning algorithm, enabling collaborative data analysis across multiple healthcare institutions while maintaining data privacy. Plaque segmentation is achieved by identifying and subtracting the External Elastic Membrane (EEM) and lumen areas. To optimize processing efficiency, polar coordinates are generated from Cartesian coordinates during the preprocessing stage, ensuring precise segmentation and improving computational performance in distributed medical imaging workflows. The model demonstrates effective plaque identification, achieving a Dice Similarity Coefficient (DSC) of 0.706, and accurately detects circular boundaries in real-time. Through collaboration with domain experts, the IVUS segmentation system provides enhanced plaque burden interpretation via precise quantitative parameter measurements. Future improvements may include integrating advanced federated learning methods to further enhance model performance. This technology is highly adaptable and suitable for environments where sensitive and distributed data are prevalent. Expanding the number of participating institutions and datasets could further optimize outcomes, enabling hospitals to achieve greater success in medical imaging and intervention.
\end{abstract}

%

\section{Introduction}\label{sec:introduction}
An invasive imaging modality such as the Intravascular Ultrasound (IVUS) image is used to evaluate Coronary Artery Diseases (CAD) and the severity of lesion stenosis \cite{CHO2024132543}. The leading cause of coronary artery diseases is atherosclerosis. Atherosclerosis is the buildup of plaque, composed of collagen, fat, calcium, cholesterol, macrophages, and microvessels \cite{STONE20202289, Saito2024}. With the accumulation of plaque and the thickening and extension of atheroma, the lumen area becomes narrow, decreasing the amount of oxygen-rich blood that could pass through. Insufficient oxygen causes coronary artery ischemia. When cardiovascular intervention with intravascular imaging is performed, images of the plaque morphology, burden, and detailed treatment strategies are provided \cite{HUANG2023102922}. Because the accumulation of plaque narrows the coronary artery, intravascular imaging is essential to estimate the plaque's location, thickness, length, and burden \cite{Li2021IVUSSeg}.

\begin{figure}[ht]
\centering
 \begin{subfigure}[c]{0.47\linewidth}
 \centering
 \includegraphics[scale=0.26]{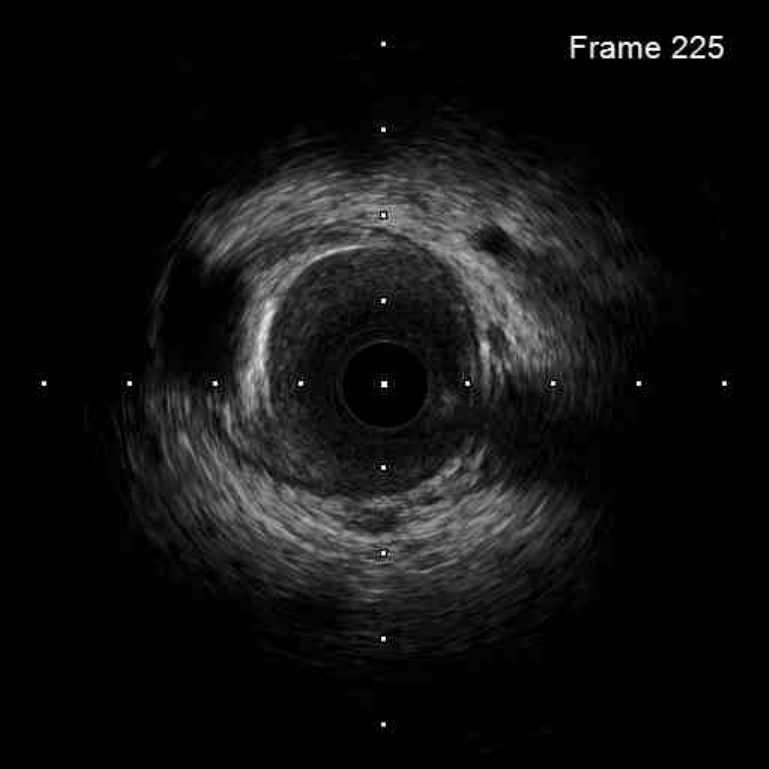}
 \caption{Original 2D IVUS image}
 \end{subfigure}\hfill
 \begin{subfigure}[c]{0.47\linewidth}
 \centering
 \includegraphics[scale=0.26]{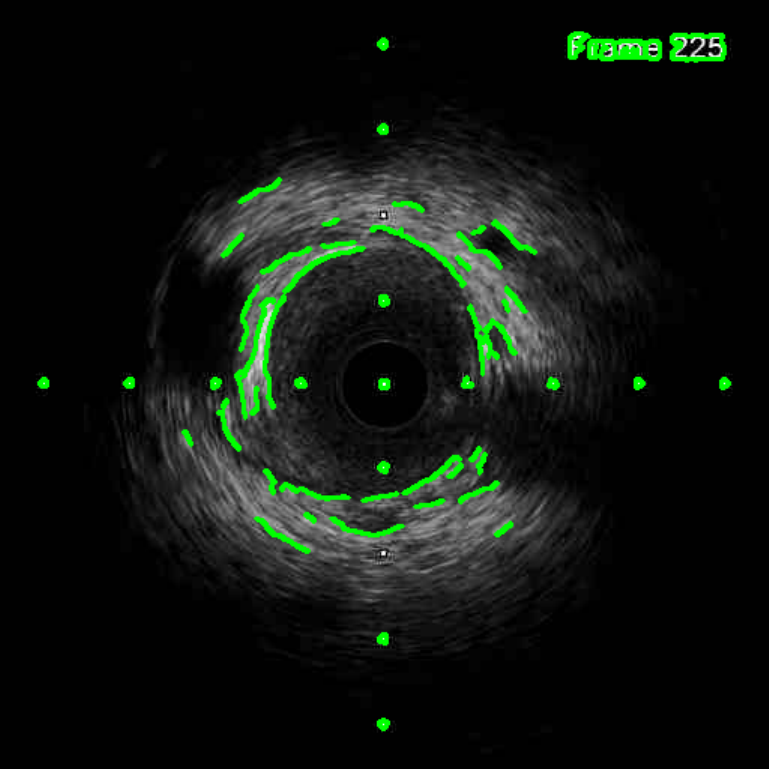}
 \caption{Result of edge detection}
 \end{subfigure}\hfill
\caption{Fragmentary boundaries after edge detection on IVUS images}
\label{fig:RelatedWork Canny Edge Detection.}
\end{figure}

IVUS provides 2D grayscale images of vessel wall structures in 360 degrees, offering real-time imaging during surgery at video-rate speed without the need for contrast, as compared to Optical Coherence Tomography (OCT) \cite{9844289, Hui2017IVUS}. Despite its advantages, IVUS generates low-resolution grayscale images with indistinct borders, as illustrated in Figure \ref{fig:RelatedWork Canny Edge Detection.}. This low resolution, common in ultrasonography, makes diagnosis more challenging. Plaque regions on IVUS images are typically identified manually by doctors using their naked eyes, relying heavily on extensive experience and professional knowledge. However, this traditional approach carries the risk of misjudgment due to the inherent subjectivity and complexity of the process.

This paper applied image segmentation technology to patients' IVUS images. The multi-stage IVUS image segmentation models mark the positions of the External Elastic Membrane (EEM), the lumen area, and the plaque, respectively. However, because of security concerns, cross-hospital medical data cannot be easily exchanged \cite{Liu_Chen_Zhao_Yu_Liu_Bao_Jiang_Nie_Xu_Yang_2022, Yan2021FL}. Therefore, the "federated learning" algorithm is proposed to address these issues and solve the cross-hospital data problem. The federated learning framework is successfully proposed to enable all cross-hospital institutions to create win-win results. It improves the feasibility of implementation. The designed model is an indispensable and effective segmentation tool in the surgical process. 

\textbf{Contribution:} Our main contributions can be summarized as follows: 
\begin{itemize}
 \item A model for automatic IVUS segmentation is proposed to highlight plaque borders and locations in arteries. The model can provide spatial information and improve image readability.
 \item In diagnosis, the proposed method is intended to overcome the difficulty of detecting circular boundaries (Figure \ref{fig:RelatedWork Canny Edge Detection.}) for improving the treatment efficiency by area segmentation models. Data are preprocessed with coordinate conversions, and lumen and EEM areas are segmented with parallel U-Net models. The plaques are identified by subtracting both locations from one another.
 \item Moreover, the proposed federated learning architecture significantly diagnoses and classifies plaques based on intravascular ultrasound images. The system gives all inter-hospital institutions a win-win situation and makes implementation easier. Additionally, it can estimate the volume of the lesion area based on the position of the external elastic membrane, lumen, and plaque. It's essential for effective segmentation during surgery.
\end{itemize} 

\begin{figure}[ht]
\centering
\includegraphics[scale=0.24]{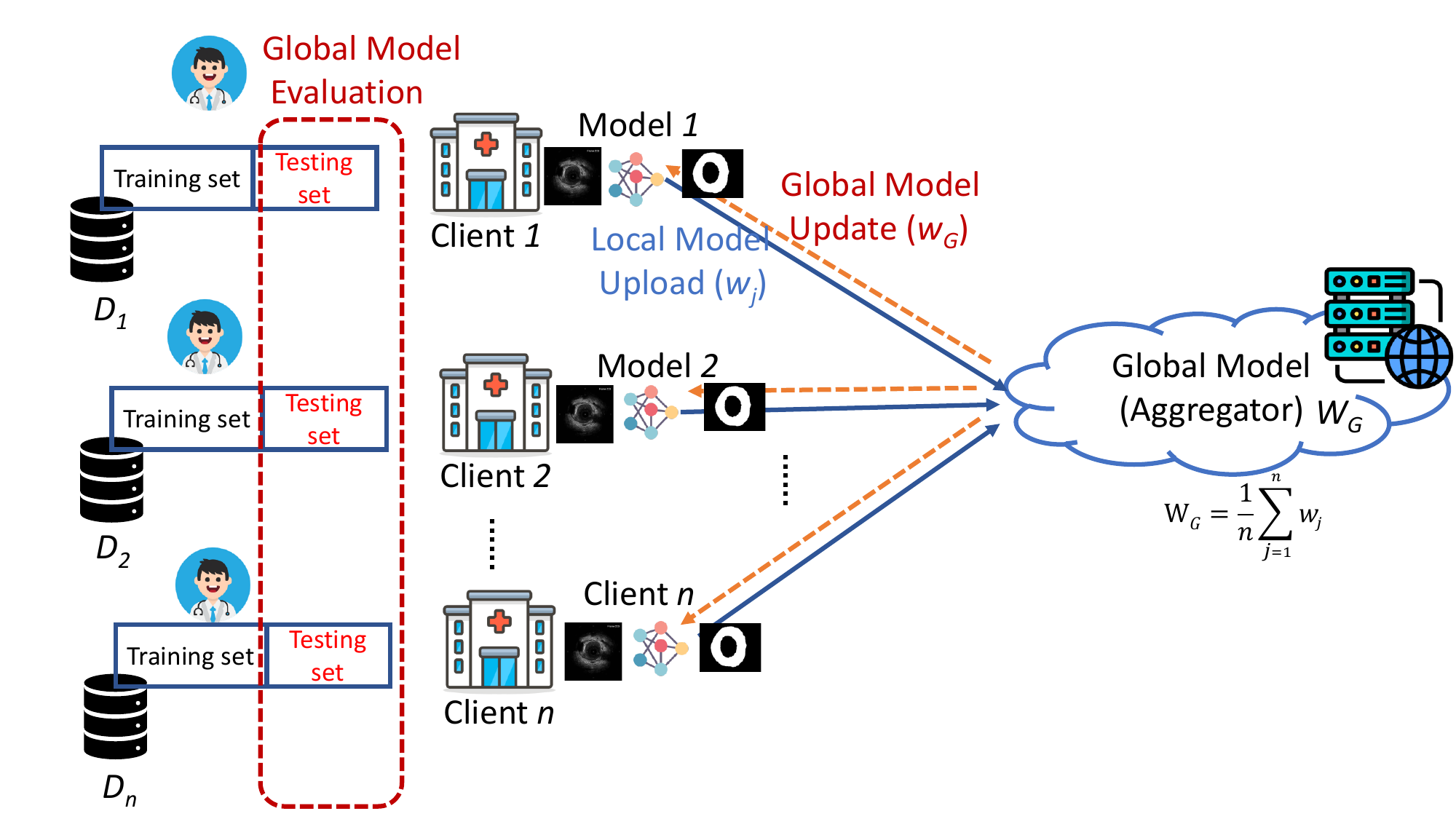}
\caption{Federated learning framework among multi-hospitals.}
\label{fig: Methods FL Procedure.}
\end{figure}

\section{Related Works}\label{sec:relatedwork}

Convolutional Neural Networks (CNNs) have been widely employed in diverse image analysis applications \cite{Saito2024}, demonstrating robust performance in numerous studies. With GPU acceleration, the high computational demands of CNNs become manageable, enabling real-time predictions. Hou \emph{et al.} \cite{Hou2020MultiStage} introduced a three-stage network for kidney tumor segmentation, utilizing a low-resolution model to locate the Volume of Interest (VOI) from down-sampled CT images \cite{MIN2022PlaqueSeverity}. Subsequently, a full-resolution model and a tumor refinement model extracted precise kidney and tumor boundaries from full-resolution CT images \cite{OPLAB2022kidney}. Similarly, Liu \emph{et al.} \cite{Liu2020Pancreas} employed fully FCNs to tackle the challenging task of pancreas segmentation in CT images.

Advancements in computational power have enabled the creation of sophisticated models that maintain high efficiency. Among these, the encoder-decoder framework U-Net has demonstrated exceptional performance in various biomedical segmentation tasks. Building on this foundation, the U-Net++ model introduces a nested structure with varying depths \cite{Zhou2020Unetplus, doi:10.1177/01617346221114137}, allowing multi-scope feature fusion. This innovative design significantly enhances segmentation accuracy, particularly in complex biomedical imaging applications.

The first federated learning algorithm was proposed in \cite{McMahan2017FedAvg}, in which a model of next-word suggestions was created in Gboard, and a collaborative learning algorithm was developed. Multiple clients can learn a global model without exchanging data \cite{Hao2020FL}. In several fields, federated learning has been demonstrated as an advantageous method for solving tasks on large amounts of data. Kumar \emph{et al.} \cite{Kumar2021FL} integrated blockchain technology with federated learning to identify confirmed COVID-19 cases using Computed Tomography (CT) images during the global spread of the pandemic. Similarly, Zhang \emph{et al.} \cite{Zhang2021FL} introduced a dynamic fusion-based federated learning approach for detecting positive COVID-19 cases. In the industrial domain, federated learning has been employed to develop robust central models while ensuring data privacy \cite{Zhao2021FL}.

\textbf{Summary:} Theoretically, federated learning can train models on multiple datasets without transferring data, but it is not easy practically. Based on recent studies, federated learning has been applied in the classification of MRI and CT images and has yielded outstanding results; however, image segmentation on IVUS images has yet to achieve significant results. Despite the importance of considering all aspects, federated learning still needs a comprehensive framework. Regarding the scarcity and sensitivity of medical data, this study proposes a lightweight deep-learning segmentation model for identifying plaques on IVUS images in real time based on a federated learning framework. 

Assuming that each local dataset has a limited volume of data and IVUS images have a low resolution, it is more difficult to identify them than MRIs and CTs. Nevertheless, the IVUS has a low resolution, and plaque is usually unclear. Therefore, a multi-stage strategy is appropriate. This study proposes a multi-stage training process for developing a lightweight IVUS segmentation model. Segmentation provides spatial information about the EEM, lumen areas, and plaques, which is very useful. In further applications, an automatic real-time segmentation model can provide valuable messages to doctors for processing IVUS images during clinical surgery.

\section{Proposed Methods}\label{sec:proposedmethods}
\subsection{Ethical Approval}
The protocol and the request for a waiver of informed consent for retrospective data collection and the use of existing biosamples (REC No. 2021-05-002B) were approved during the 136th meeting of the Institutional Review Board of Taipei Veterans General Hospital on May 14, 2021. In accordance with Good Clinical Practice guidelines and applicable government laws and regulations, all experiments were conducted under the oversight of the committee. Written informed consent from participants was not required for this study.

A federated learning framework is proposed for plaque segmentation on distributed IVUS datasets, utilizing self-designed U-Net models. The procedures of the federated learning framework are illustrated in Figure \ref{fig: Methods FL Procedure.}. In this framework, local models are aggregated into a global model ($W_G$) using the FedAvg algorithm \cite{McMahan2017FedAvg}. The client repeatedly exchanges model weights ($W_j$) with the server, a process referred to as "communication." The framework is tested with \textit{N} clients, where \textit{N} represents the number of participants in the federated learning environment. This study involves a limited number of participating clients ($N=3$). Local clients implement the proposed U-Net architecture with consistent hyperparameter settings across the federated learning framework. 

\begin{equation}
\label{eqn: Dice Score}
 w_{G} \leftarrow \sum_{j=1}^{N}{\frac{D_{j}}{D} \ w_{j}},
\end{equation}



where $D$ represents the total dataset, while $D_{j}$ denotes the proportion of data held by client $j$ relative to the total dataset ($D$). In Algorithm \ref{alg: FedAVG}, the global model is denoted by $w_{G}$, and the model weights of each client are represented by $w_{j}$. The server aggregates all received parameters from clients to generate an updated global model, $w_{G}^{r+1}$, as the iteration $r$ increases. Once the server model is updated, it is broadcast to all local clients, instructing them to retrain using their local datasets ($D_j$). After completing local training, each client sends their updated model parameters ($w_{j}^{r+1}$) back to the server. This process continues until the training program concludes or no further performance improvements are observed.


\newcommand{\algrule}[1][.2pt]{\vskip.5\baselineskip\hrule height #1\vskip.5\baselineskip}
\algnewcommand\algorithmicinparalleldo{\textbf{in parallel do}}
\algdef{S}[FOR]{ForParallel}[1]{\algorithmicfor\ #1\ \algorithmicinparalleldo}

\begin{algorithm}[ht]
\small
\caption{FedAvg \cite{McMahan2017FedAvg}}
\label{alg: FedAVG}

 \begin{algorithmic}[1]
 \Require{
 Number of communication rounds $R$,\newline
 Number of local epochs $E$,\newline
 Local minibatch size $B$,\newline
 Learning rate $\eta$
 }
 \Ensure{Global Model $W_G$}
 \algrule
 
 \Function{SERVERPROCESS}{}\Comment{Run on the server}
 \State initialize $w_{0}$
 \State $S$ $\leftarrow$ (a set of $N$ clients)
 \For{each round $r$ from $1$ to $R$}
 \ForParallel{each client $j \in S$}
 \State $w_{j}^{r+1} \leftarrow \text{CLIENTUPDATE}(j, w, E)$
 \EndFor
 \State $w_{G}^{r+1} \leftarrow \sum_{j \in S}{\frac{D_{j}}{D} \ w_{j}^{r+1}}$
 \EndFor
 \EndFunction
 \\
 \Function{CLIENTUPDATE}{$j, w, E$}\Comment{Run on client $j$}
 \State $\mathcal{B} \leftarrow$ (split $D_j$ into batches of size $B$)
 \For{each local epoch $e$ from $1$ to $E$}
 \For{batch $b \in \mathcal{B}$}
 \State $w \leftarrow w - \eta \triangledown l(w,b)$
 \EndFor
 \EndFor
 \State return $w$ to server
 \EndFunction
 \end{algorithmic}
\end{algorithm}

\subsection{Parallel Model Architecture and Data Coordinate Conversion}
\begin{figure}[ht]
\centering
\includegraphics[scale=0.26]{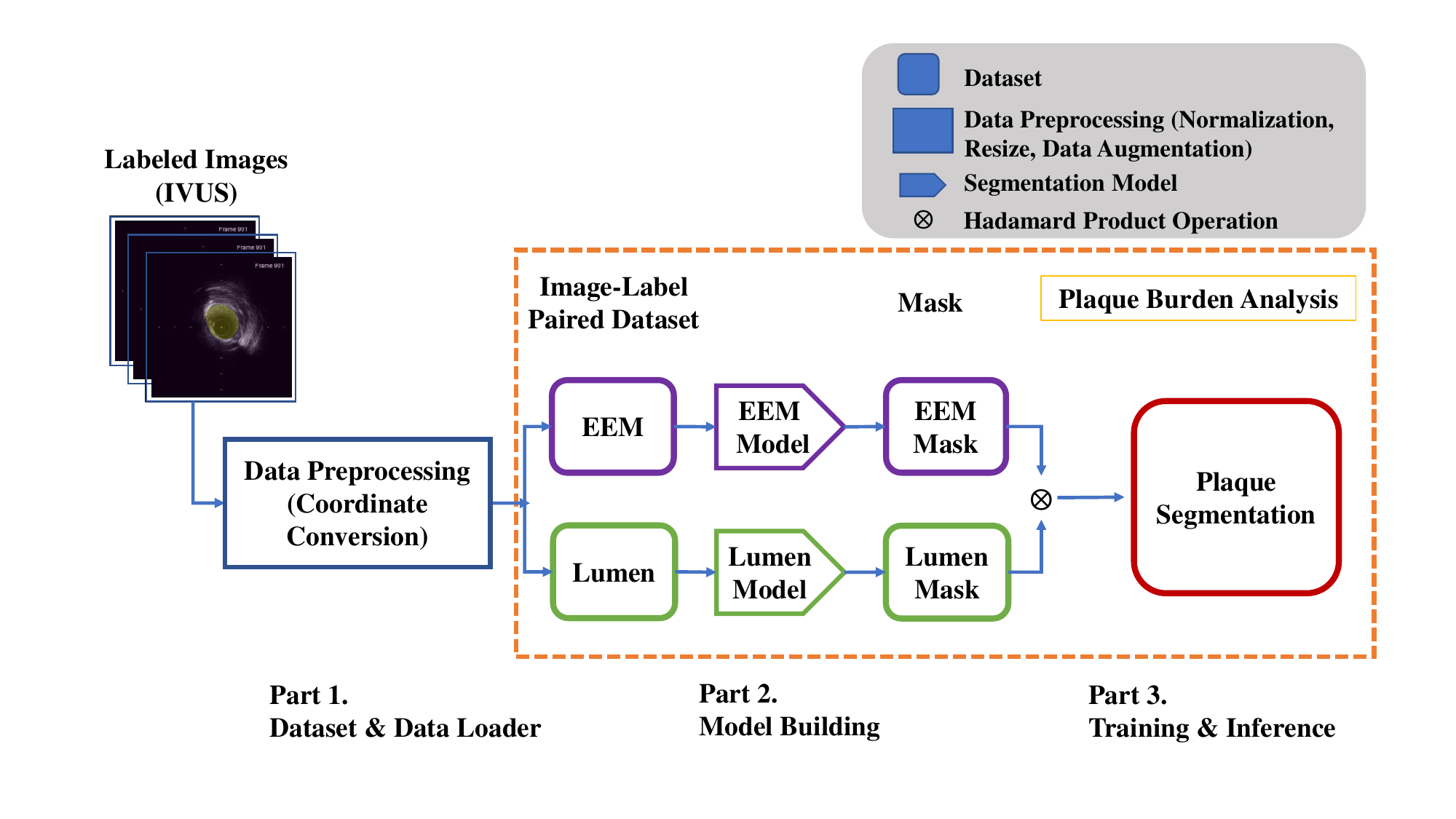}
\caption{Proposed parallel plaque segmentation diagram. }
\label{fig: Methods system process.}
\end{figure}

Physicians locate the plaque burden area within the cardiovascular by observing the EEM border and lumen border because plaque grows between those two areas. To simulate the diagnostic process, the plaque segmentation task is divided into two stages. The first step generates segmentation masks for each input image by identifying the EEM and lumen areas. The next step involves subtracting the segmentation masks of the EEM and lumen areas to form a plaque mask. A two-stage procedure is performed sequentially to determine the plaque burden. Thus, the federated learning framework is used to train the model. In each hospital, the diagram shows the overall process and details (Figure \ref{fig: Methods system process.}).  

Coordinate systems for plane images include Cartesian and polar systems. Cartesian coordinates, being the most intuitive for human interpretation, allow for straightforward observation of spatial relationships and preserve object shapes, making them highly effective for image analysis. Plaque detection is commonly performed using Cartesian coordinate images due to their ease of interpretation, where each pixel is defined by two indices, such as \textit{(x, y)}. In this study, IVUS frames were extracted from the Digital Imaging and Communications in Medicine (DICOM) raw cross-sectional view in Cartesian coordinates. All images and masks utilized in the proposed deep learning model are formatted in a 512×512 shape, with both inputs and outputs processed in Cartesian coordinates. This ensures consistency and compatibility throughout the analysis pipeline.

In polar coordinates, images present differently than in Cartesian coordinates. Each pixel is defined by its distance from a reference point and its angle relative to a reference direction. Cross-sectional views of vessels appear approximately circular in this coordinate system. Cartesian coordinate images and masks can be transformed into polar coordinates to incorporate angular information \cite{CHO2021polar}. Assuming the center of the original image as the reference point, the radius of the largest circle is set to 256 pixels, and polar coordinates are sampled every 0.5 degrees for precision. In the resulting polar images, the horizontal axis represents the angle ($\theta$), and the vertical axis denotes the distance ($r$) from the pixel to the center. This transformation produces images sized 256×720 pixels, approximately 70\% of the original size.

\begin{figure}[ht]
\centering
\includegraphics[scale=0.27]{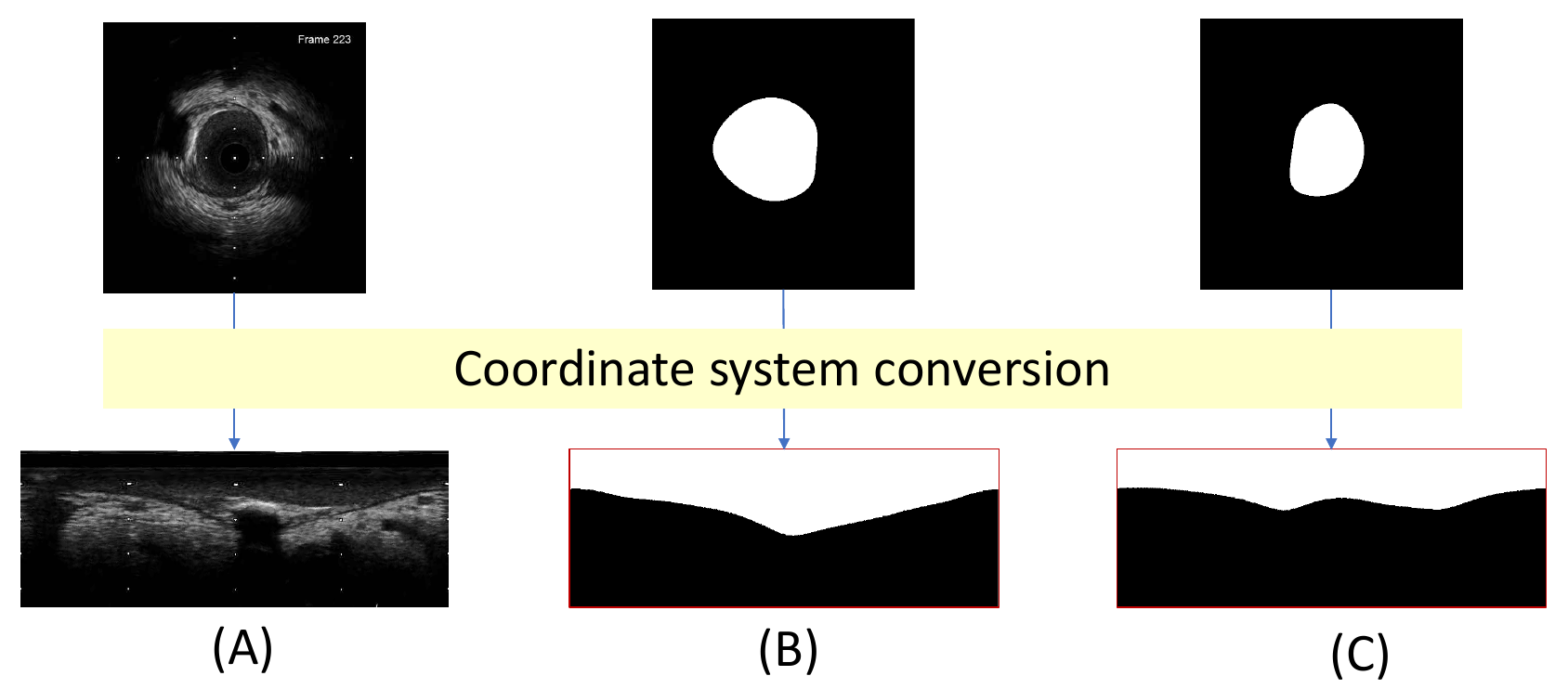}
\caption{(A) Original 2D IVUS image (B) EEM mask (C) lumen mask in polar coordinate system.}
\label{fig: IVUS images in polar coordinate system.}
\end{figure}

As shown by Cho \emph{et al.} \cite{CHO2021polar}, converting IVUS images from Cartesian to polar coordinates allows for angle-wise plaque classification, enabling accurate determination of plaque orientation. Medical images were collected in the DICOM format, representing cardiovascular structures in three dimensions. This study employs two-dimensional frames represented in both Cartesian and polar coordinate systems. Figure \ref{fig: IVUS images in polar coordinate system.} illustrates an IVUS frame, its EEM mask, and its lumen mask converted to polar coordinates. The transformation filters out the background from the corners of the original image and reveals how boundaries change when viewed from different angles in the converted images.

\section{Experiments}\label{sec:experiments}
\subsection{Implementation Parameter and Environment}
This study employs a deep learning model implemented using Python 3.9.2 and the open-source machine learning framework PyTorch v1.8.0. Computational processes are accelerated with an Nvidia GeForce RTX 2070 Super graphics card featuring an 8 GB frame buffer. The model training utilizes a hybrid loss function that combines Cinary Cross-Entropy Loss (BCELoss) and Dice Similarity Coefficient Loss (DSCLoss). The DSC value measures the similarity between two images. To calculate the values between the DSC, use Equation \ref{eqn_DiceScore}, where the annotated and output segmentation masks are denoted by \textit{A} and \textit{B}. This formula represents twice the intersection of the Region of Interest (ROI) compared to the sum of the ROIs in the two masks. The coefficient is 1 when the segmentation mask perfectly matches the ground truth mask; however, it becomes 0 if it does not match any pixels. According to the above definition, the Dice similarity coefficient is precisely the harmonic mean of precision and recall, which can also be expressed as Equation \ref{eqn_DiceScoreHarmonic}. The segmentation model is trained using a combined loss function. This loss function combines the BCELoss and DSCLoss see Equation \ref{eqn: Hybrid loss function}. A segmented image can be regarded as a classification of each pixel. The parameters $\omega$ and $1 - \omega$ are weights of the above two terms in our loss function, and both are set to 0.5.

\begin{equation}
\label{eqn_DiceScore}
 \text{DSC}=\frac{2\times |A\cap B|}{|A|+|B|}
\end{equation}

\begin{equation}
\label{eqn_DiceScoreHarmonic}
 \text{DSC} = 2 \times \frac{\text{Recall} \times \text{Precision}}{\text{Recall} + \text{Precision}}
\end{equation}

\begin{equation}
\label{eqn: Hybrid loss function}
 \text{Loss value}= \omega \times \text{BCELoss} + (1 - \omega) \times \text{DSCLoss}
\end{equation}

Model updates are performed using the Adaptive Moment Estimation (Adam) optimizer with a learning rate of 1.00e-5, and L2 regularization is applied to prevent overfitting by incorporating a regularization term. Due to computational constraints, all experiments are conducted with a batch size of 4. To evaluate model performance, five-fold cross-validation is employed, partitioning the dataset into five independent folds at the patient level, ensuring a balanced distribution of patients. For the federated learning setup, the dataset is divided into three folds, each assigned to a local client (Fig. \ref{fig: Methods FL Procedure.}). Model training is conducted for 10 communication rounds, with each local client performing 1 epoch of training per round. This setup enables collaborative construction of a global model across distributed datasets.

\subsection{Dataset Description}
\begin{figure}[ht]
\centering
\includegraphics[scale=0.25]{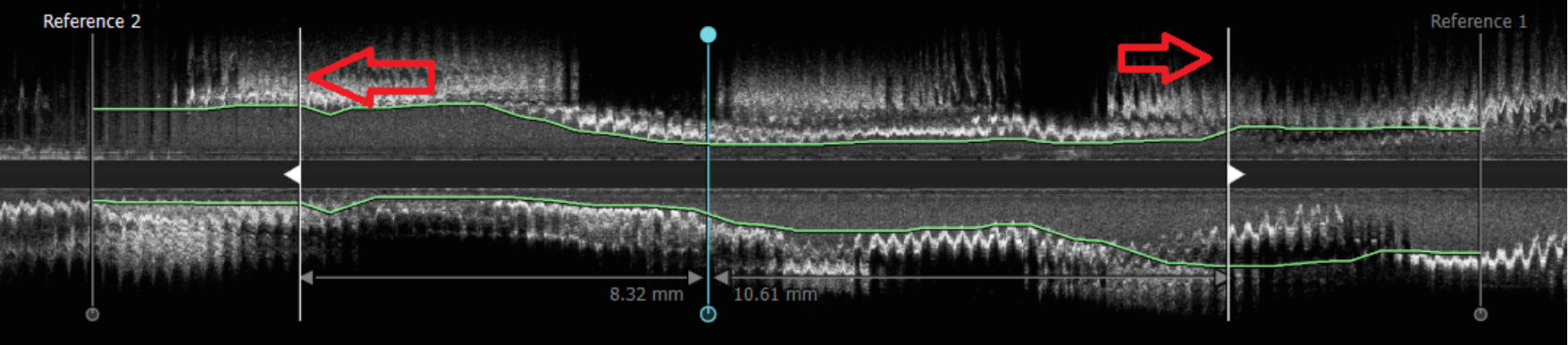}
\caption{Cardiovascular appearance in the longitudinal view.}
\label{fig: cardiovascular in the longitudinal view.}
\end{figure}

The IVUS dataset (IRB: 2021-05-002B) was collected from three independent hospitals, Taipei Veterans General Hospital (TVGH), Taichung Veterans General Hospital (VGHTC), and Kaohsiung Veterans General Hospital (VGHKS), Taiwan. A total of 151 individuals that undergo PCI and IVUS were included in the study. An IVUS mechanical system (OptiCross HD, 60 MHz Coronary Imaging Catheters, Boston Scientific Corporation) equipped with a 60 MHz wideband transducer collects raw data in DICOM format. Two physicians blinded to the participant information annotated approximately 87,505 frames of 2-dimensional images from 151 individuals. A cardiovascular is annotated within the target length indicated by the white line in the longitudinal view, which is the narrowing range of the lumen area owing to plaque growth as shown in Figure \ref{fig: cardiovascular in the longitudinal view.}. Each adjacent frame is approximately 3 mm apart. There are a certain number of frames depending on the length of the cardiovascular system. There are two classes in total, and each frame is annotated at the pixel level; class "0" for the non-region-of-interest area and class "1" for the region-of-interest area, which include the EEM, lumen, or plaque burden areas.

To assess the inter-observer variability for the EEM, lumen, and plaque burden area, two independent domain experts (doctors or cardiologists), evaluated 3,000 randomly selected images. In Table \ref{tab: Inter-observer agreement}, the DSC is calculated for the inter-observer image analysis. The consensus between the two analysts determines the final annotations, and a senior expert resolves conflicts. Despite the conflicting opinions, there was an excellent inter-observer agreement between the two cardiologists for the EEM area annotations (DSC $0.753$), lumen area annotations (DSC $0.793$), and plaque burden index annotations (DSC $0.778$).

\begin{table}[ht]
 \newcommand{\tabincell}[2]{\begin{tabular}{@{}#1@{}}#2\end{tabular}}
 \caption{Comparison of Segmentation Results for EEM, Lumen, and Plaque Burden Between the Proposed Model and Expert Annotations (DSC).}
 \label{tab: Inter-observer agreement}
 \centering
 \renewcommand{\arraystretch}{1.3}
 \begin{tabular}{p{6em}p{3.5em}p{3.5em}p{6em}}
  \hline\noalign{\smallskip}
  \bfseries \tabincell{l}{} &
  \bfseries \tabincell{c} EEM &
  \bfseries \tabincell{c} Lumen &
  \bfseries \tabincell{c} Plaque Burden Index\\
  \hline\noalign{\smallskip}
  Expert Annotations & {0.753} & {0.793} & {0.778}\\
  Proposed Approach & {0.890} & {0.877} & {0.706}\\
  \hline
 \end{tabular}
\end{table}

\subsection{Experimental Results and Analyses}
Plaque in the cardiovascular system can obstruct ultrasound reflections, causing blood vessels to appear unclear on IVUS images. This results in manually annotating IVUS images becoming a time-consuming and labor-intensive process. Analysts must often rely on their expertise and experience to infer boundaries that are unclear or disconnected. Even among highly trained professionals, interpretations of the same image may vary, leading to non-identical annotations. According to Table \ref{tab: Inter-observer agreement}, two independent domain experts evaluated the same dataset, achieving Dice similarity coefficients of approximately 0.778 for the plaque burden index and 0.753 to 0.793 for the EEM and lumen. These results demonstrate excellent inter-observer agreement, highlighting the consistency of expert evaluations despite the inherent challenges.

Quantitative indices demonstrate Dice similarity coefficients for segmenting the EEM, lumen, and plaque burden. The proposed parallel deep learning models were evaluated using five-fold cross-validation to segment these structures across different image coordinates. Key performance metrics, including Dice similarity coefficient, recall, and precision, were assessed. For EEM and lumen segmentation with coordinate conversion processing, the model achieved metrics exceeding 0.877, indicating high accuracy. Conversely, plaque segmentation resulted in metrics around 0.706, closely matching the results of domain experts and highlighting the increased complexity involved in accurately identifying plaque boundaries.

Model inference time was also considered, emphasizing the need for real-time IVUS image segmentation during surgical procedures. An efficient model not only supports real-time operations but also enhances the framework's overall efficiency for deep learning tasks and federated learning applications, demonstrating its versatility and practical utility.

\subsubsection{Plaque Segmentation Results:} 
\begin{figure}[ht]
\centering
\includegraphics[scale=0.20]{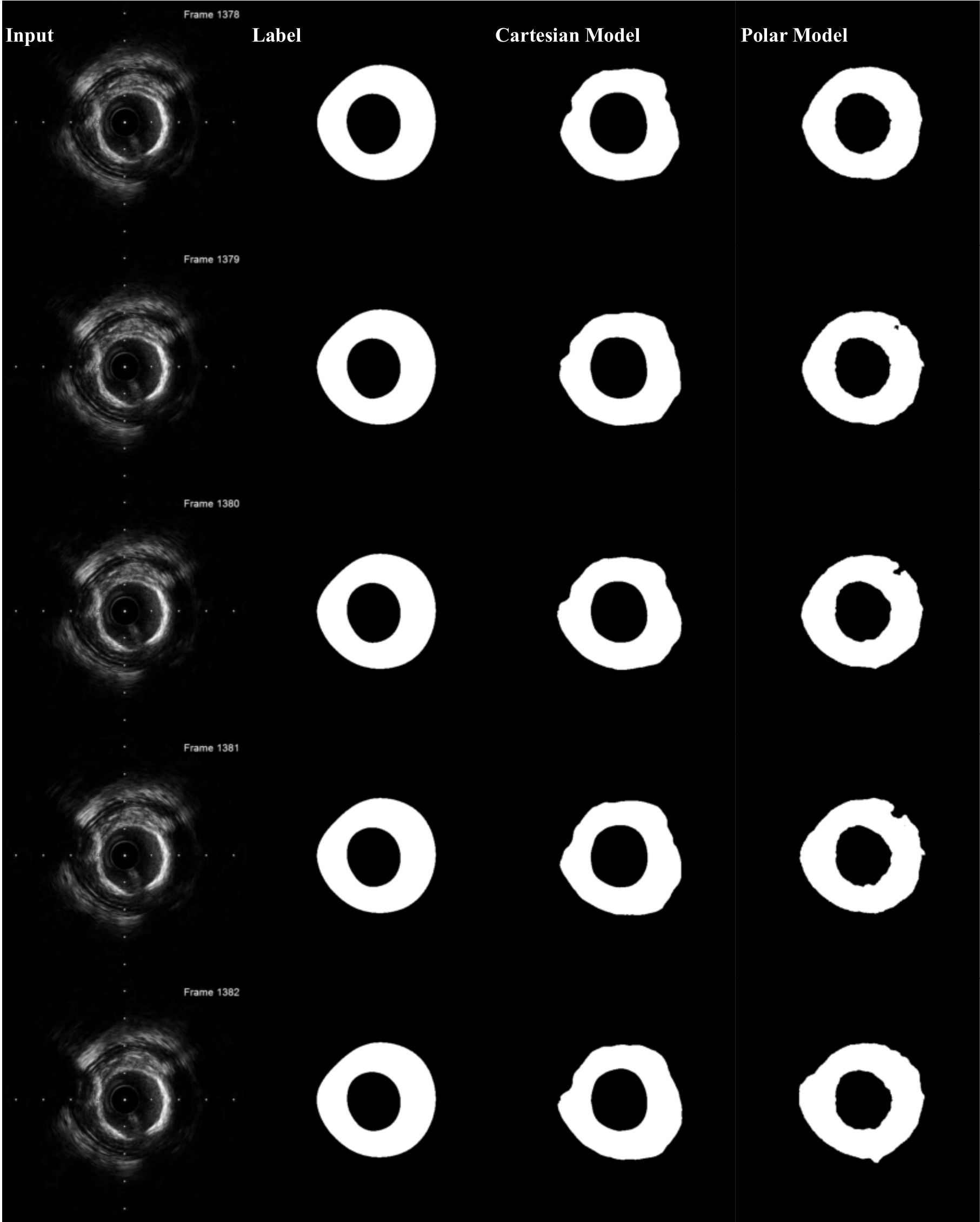}
\caption{Segmentation results of Cartesian system and polar coordinate conversion with post-processing.}
\label{fig: SegmentationResults_Comparison}
\end{figure}

Based on the quantitative analysis presented earlier, image segmentation in polar coordinates with post-processing improves the EEM segmentation results, particularly enhancing plaque burden segmentation. Detecting EEM areas is often challenging due to plaque burden artifacts that result in disconnected borders. A comparison of the plaque burden segmentation outcomes between the two methods is illustrated in Figure \ref{fig: SegmentationResults_Comparison}.

The first and second columns display five consecutive frames along with their corresponding annotation masks. The third and last columns show the segmentation results obtained using Cartesian and polar coordinate methods, respectively. Both methods demonstrate comparable performance for these frames, achieving DSC scores of approximately 0.89. However, the Cartesian coordinate method exhibits notable signal loss on the right and left sides of the IVUS image, particularly in the fourth quadrant. This results in indentations on the left and right sides of the predicted segmentation outcomes. Artifacts result in boundaries that appear concave inward. While the polar coordinate method may produce jagged edges, the contours remain sharp and well-defined, minimizing the impact of such artifacts. Consequently, segmentation using polar coordinates delivers superior and more accurate results compared to other methods.

\subsubsection{Consistency Evaluation Between Manual and Automatic Measurements:} The analyzed indicators include the EEM area, lumen area, plaque burden area, plaque burden index, EEM volume, lumen volume, and plaque burden volume. The plaque burden index, as defined by Equation \ref{eqn: Plaque Burden Index}, represents the proportion of the plaque burden relative to the EEM area. To evaluate the agreement between manual measurements and automatic estimations by the global model, a Bland-Altman plot, also known as a Tukey mean difference plot, is utilized.

\begin{equation}
\label{eqn: Plaque Burden Index}
 \begin{aligned}
 \text{Plaque Burden Index} &= \frac{\text{EEM Area} - \text{Lumen Area}}{\text{EEM Area}}\\\\
 &= \frac{\text{Plaque Area}}{\text{EEM Area}}\\
 \end{aligned}
\end{equation}\\

This study utilized 87,505 IVUS frames from 151 patients collected across three hospitals (TVGH, VGHTC, and VGHKS). The dataset was divided into 135 cases (78,749 frames) for training and 16 cases (8,756 frames) for global verification. A single patient could contribute multiple frames, resulting in a non-independent dataset.
The plaque burden index, defined by the EEM, lumen, and plaque areas as shown in Equation \ref{eqn: Plaque Burden Index}, quantifies vascular blockage. Lower index values complicate plaque identification and influence training due to data distribution variations. Risk thresholds of 50\% and 70\% classify cardiovascular disease risk, with indices below 50\% indicating low risk and above 70\% signifying high risk \cite{10.1001/jamacardio.2023.2731}. Table \ref{table: description N=3 IID} summarizes dataset characteristics for federated learning with balanced and similar distributions.

\begin{figure*}[ht]
\centering
\begin{multicols}{3}
 \begin{subfigure}{0.32\textwidth}
 \centering
 \includegraphics[width=1\linewidth]{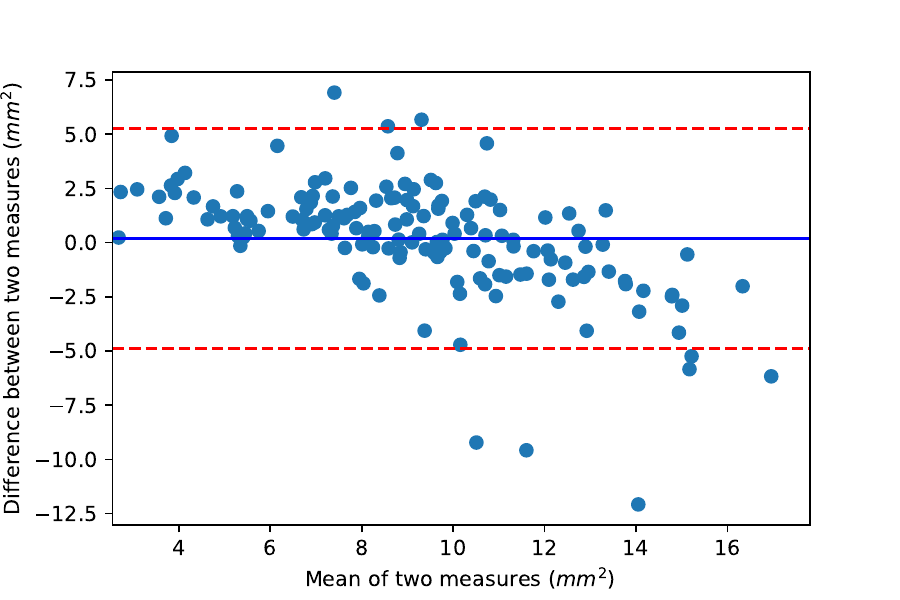}
 \caption{Plaque Area}
 \label{subfig: polar Plaque area BA}
 \end{subfigure}
 \begin{subfigure}{0.32\textwidth}
 \centering
 \includegraphics[width=1\linewidth]{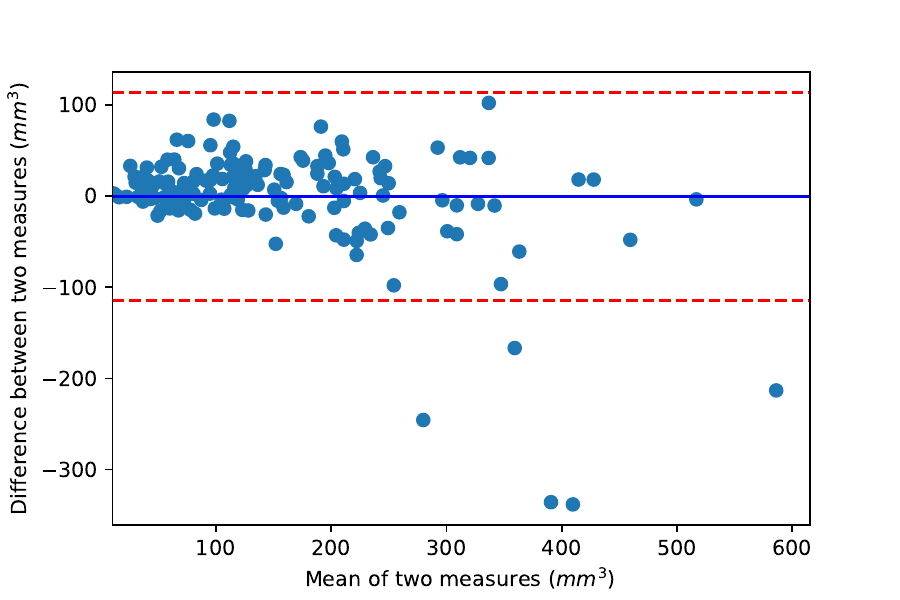}
 \caption{Plaque Volume}
 \label{subfig: polar Plaque volume BA}
 \end{subfigure}
 \begin{subfigure}{0.32\textwidth}
 \centering
 \includegraphics[width=1\linewidth]{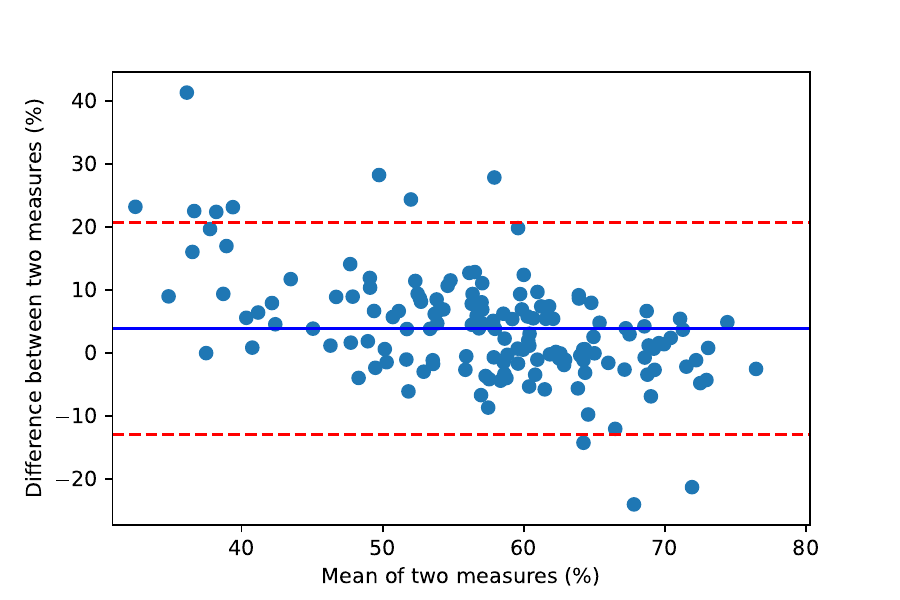}
 \caption{Plaque Burden Index}
 \label{subfig: polar Plaque Burden Index BA}
 \end{subfigure}
\end{multicols}
\caption{Bland-Altman plot of quantitative indicators for plaque estimation comparing the global model with the polar coordinate conversion to domain experts' measurements.}
\label{figpolarBAPlot}
\end{figure*}

\begin{figure*}[ht]
\centering
\begin{multicols}{3}
 \begin{subfigure}{0.33\textwidth}
 \centering
 \includegraphics[width=1\linewidth]{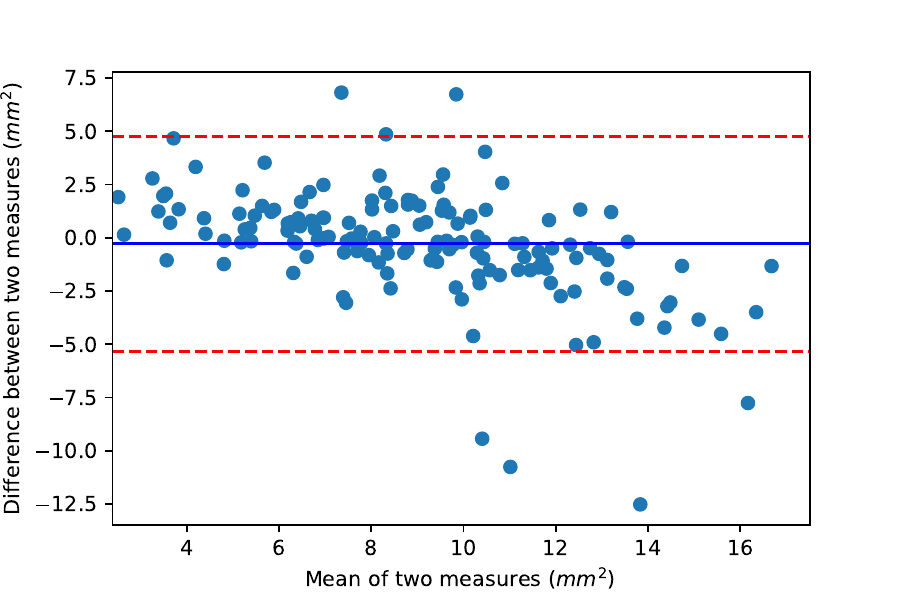}
 \caption{Plaque Area}
 \label{subfig: cart Plaque area BA}
 \end{subfigure}
 \begin{subfigure}{0.33\textwidth}
 \centering
 \includegraphics[width=1\linewidth]{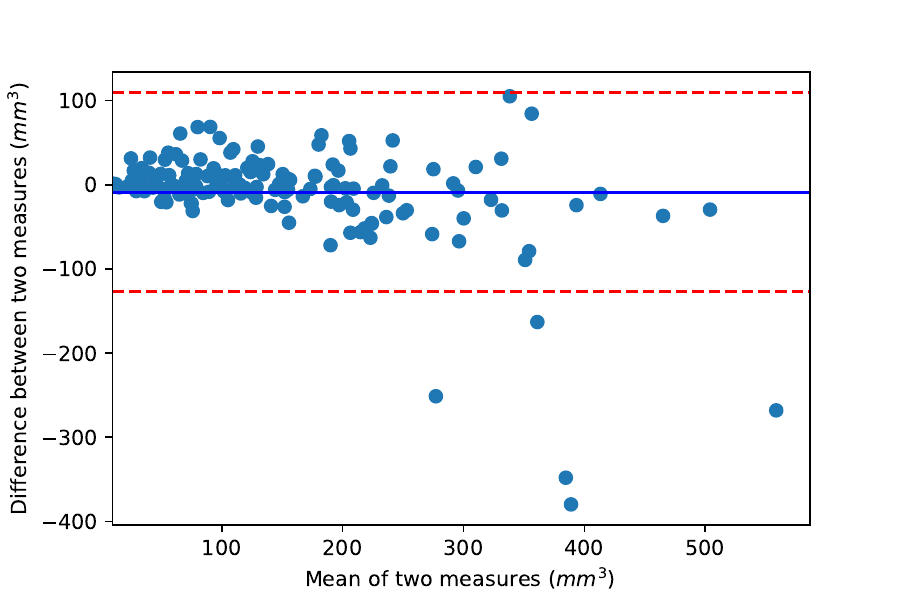}
 \caption{Plaque Volume}
 \label{subfig: cart Plaque volume BA}
 \end{subfigure}
 \begin{subfigure}{0.33\textwidth}
 \centering
 \includegraphics[width=1\linewidth]{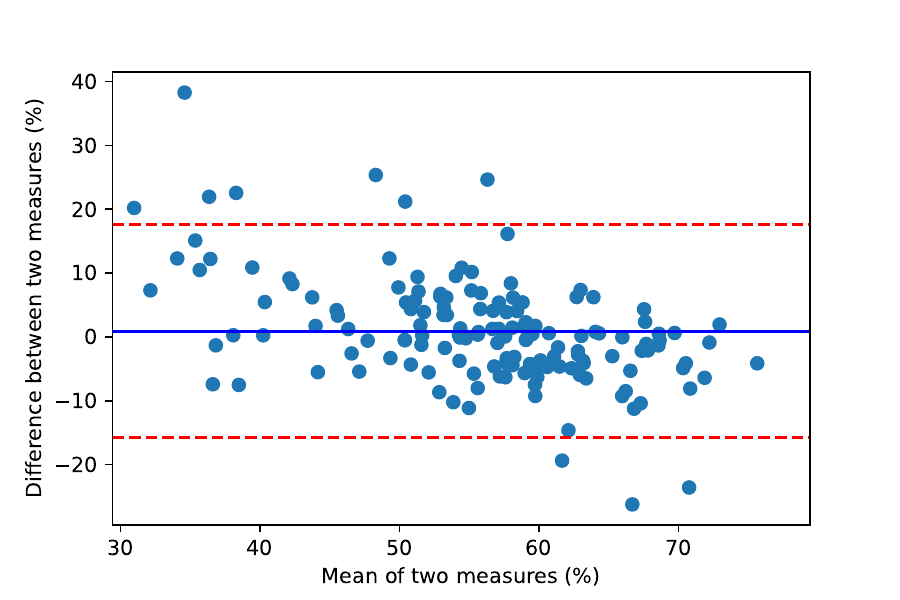}
 \caption{Plaque Burden Index}
 \label{subfig: cart Plaque Burden Index BA}
 \end{subfigure}
\end{multicols}
\caption{Bland-Altman plot of quantitative indicators for plaque estimation comparing the global model with Cartesian coordinate conversion to domain experts' measurements.}
\label{fig_cart BA-Plot}
\end{figure*}

Figures \ref{figpolarBAPlot} and \ref{fig_cart BA-Plot} present the Bland-Altman analysis comparing quantitative indicators estimated by the global model, utilizing Cartesian and polar coordinate conversions, with evaluations from domain experts, including doctors and radiologists. Each data point represents a patient case, with the horizontal axis showing the average of two measurements and the vertical axis depicting their difference. A blue line represents the mean difference, while red lines indicate the 1.96 standard deviations from the mean, forming the 95\% confidence interval. Figure \ref{fig_cart BA-Plot} shows the comparison between the Cartesian segmentation method and expert annotations, whereas Figure \ref{figpolarBAPlot} illustrates the results for the polar segmentation method. 

All indicators demonstrate mean differences close to zero, suggesting minimal bias. Most data points lie within the confidence intervals and show no discernible patterns, indicating strong agreement between the methods. However, in several subplots, data points below the blue line represent negative differences, while those above indicate positive differences caused by discrepancies between predictions and expert measurements. A few data points fall outside the confidence intervals, potentially due to imbalanced data collection. The dataset contains relatively few extreme cases with high or low indicator values, causing the deep learning models to predict values closer to the dataset's mean. To overcome cross-hospital data sharing challenges, a federated learning algorithm was introduced, enabling collaborative model development while maintaining data privacy. This framework delivers mutual benefits to participating institutions and enhances feasibility. The proposed model is an effective and indispensable segmentation tool, essential for surgical procedures.

\begin{table}[ht]
\centering
\caption{Descriptions of local datasets (balanced datasets with similar data distributions).}
\label{table: description N=3 IID}
\renewcommand{\arraystretch}{1.3}
 \begin{tabular}{
>{\centering\arraybackslash}p{2.8cm}
>{\centering\arraybackslash}p{1.2cm}
>{\centering\arraybackslash}p{1.2cm}
>{\centering\arraybackslash}p{1.2cm}}
 \multicolumn{4}{r}{(\# of Frames / \# of Cases)}\\
\hline
 \textbf{Plaque Burden Index Distribution} & \textbf{Dataset 1}& \textbf{Dataset 2} & \textbf{Dataset 3}\\
\hline
 Low ($<$ 50) & \phantom{0}6,039 / 10 & \phantom{0}7,907 / 15 & \phantom{0}7,864 / 14\\
 Moderate (50 $\sim$ 70) & 16,592 / 31 & 15,737 / 27 & 15,851 / 26\\
 High ($>$ 70) & \phantom{0}3,411 / \phantom{0}4 & \phantom{0}2,714 / \phantom{0}3 & \phantom{0}2,634 / \phantom{0}5\\
 (Total) & 26,042 / 45 & 26,358 / 45 & 26,349 / 45\\
\hline
 \\\\
\end{tabular}
\end{table}

\section{Discussion}\label{sec:discussion}
IVUS provides low-resolution grayscale images with vague borders, making diagnosis particularly challenging. For junior doctors, identifying plaque regions based solely on professional experience and knowledge can be difficult. Additionally, data collection and analysis across multi-hospital systems face significant regulatory and privacy challenges, necessitating the development of algorithms with decentralized architectures.

This study introduces a two-stage deep-learning method for automatic plaque segmentation on IVUS images within a federated learning framework. The segmentation process involves converting images from Cartesian to polar coordinates, achieving Dice similarity coefficients of approximately 0.890 for EEM, 0.877 for lumen, and 0.706 for plaque burden areas. Preprocessing includes coordinate conversions, and segmentation of the lumen and EEM areas is performed using parallel U-Net models. Plaque regions are determined by subtracting the segmented locations of the lumen and EEM. This approach significantly enhances spatial information and improves image readability.

The smaller size of the converted polar images reduces the number of model parameters from approximately 933k to 721k, optimizing both parameter efficiency and input dimensions while utilizing fewer computing resources. Although post-processing slightly increases computational time due to coordinate conversion, the polar coordinate method delivers superior segmentation performance.

Additionally, this study employs a federated learning framework to segment 60-MHz IVUS images. The proposed area segmentation model improves treatment effectiveness and efficiency by alleviating the challenges of detecting circular boundaries. Plaque borders and locations within the arteries are automatically highlighted, providing valuable insights for clinical decision-making.

The limitations of this study are briefly outlined. Each local dataset contains a limited volume of data, and some IVUS images lack high resolution, with most plaques appearing unclear. The federated learning framework was tested on only three clients due to the limited number of distributed datasets. Furthermore, all clients had equal amounts and similar distributions of IVUS images, whereas real-world scenarios often involve unbalanced datasets with distinct distributions across clients. The lower resolution of 20MHz or 40MHz IVUS images further complicates patient identification across different clinics. Another significant challenge is the segmentation of images for model training, which requires a substantial number of annotated images, a considerable obstacle for federated learning frameworks.

\section{Conclusion}\label{sec:conclusion}
This study highlights practical methodologies and case studies on the IVUS segmentation system and federated learning framework, demonstrating significant real-world medical applications that are adaptable and scalable within decentralized AI systems: 1) Cardiovascular Disease Diagnosis and Treatment: The system accurately identifies and segments arterial structures, including the external elastic membrane, lumen, and plaque borders. This aids in diagnosing atherosclerosis and assessing plaque stability, which are essential for preventing heart attacks and strokes. 2) Surgical Planning: Precise segmentation provides interventional cardiologists with accurate spatial and volumetric data, supporting procedures such as stent placement and angioplasty. 3) Cross-Hospital Collaboration: Utilizing federated learning, hospitals can collaboratively train models on IVUS datasets without sharing sensitive patient data. This approach facilitates multi-center studies and ensures model generalization across diverse populations. 4) Enhanced Image Analysis: Enhanced image readability and spatial information allow clinicians to interpret IVUS images more accurately, reducing diagnostic errors and improving patient outcomes, even when working with a limited number of imaging datasets. 5) Personalized Medicine: The integration of demographic and patient-specific data enables the creation of personalized cardiovascular disease risk profiles, allowing for tailored treatment plans and preventive strategies. 6) Broader Medical Applications: The federated learning framework is adaptable to other imaging modalities, such as CT, MRI, and ultrasound. This expands its utility to fields like oncology, neurology, and prenatal care, particularly where sensitive and distributed data are prevalent. The focus remains on delivering practical, actionable solutions ready for real-world implementation.

\section{Acknowledgments}
This work was partially supported by Academia Sinica and the National Science and Technology Council, Taiwan, under Grants 3012-C3901, 113-2221-E-002-183-MY2, 113-2321-B-075-006, and 112-2221-E-001-024-MY2. The contributions of the participants to this research are also deeply acknowledged, including Prof. Frank Yeong-Sung Lin, Prof. Yennun Huang, Mr. Tzu-Lung Sun, Dr. Wen-Lieng Lee, Dr. Feng-Yu Kuo, Dr. Tse-Min Lu, and others. Their support and collaboration have been invaluable to the success of this study.



\bibliography{aaai25}

\begin{thebibliography}{22}
\providecommand{\natexlab}[1]{#1}

\bibitem[{Cho et~al.(2021)Cho, Kang, Min, Lee, Kim, Kang, Kang, Lee, Ahn, Park, Lee, Kim, Lee, Park, and Park}]{CHO2021polar}
Cho, H.; Kang, S.-J.; Min, H.-S.; Lee, J.-G.; Kim, W.-J.; Kang, S.~H.; Kang, D.-Y.; Lee, P.~H.; Ahn, J.-M.; Park, D.-W.; Lee, S.-W.; Kim, Y.-H.; Lee, C.~W.; Park, S.-W.; and Park, S.-J. 2021.
\newblock Intravascular ultrasound-based deep learning for plaque characterization in coronary artery disease.
\newblock \emph{Atherosclerosis}, 324: 69--75.

\bibitem[{Cho et~al.(2024)Cho, Cho, Min, Lee, Kim, Lee, Lee, and Kang}]{CHO2024132543}
Cho, S.; Cho, H.; Min, H.; Lee, J.-G.; Kim, T.~O.; Lee, P.~H.; Lee, S.-W.; and Kang, S.-J. 2024.
\newblock Clinical impact of deep learning-derived intravascular ultrasound characteristics in patients with deferred coronary artery.
\newblock \emph{International Journal of Cardiology}, 417: 132543.

\bibitem[{Hao et~al.(2020)Hao, Li, Luo, Xu, Yang, and Liu}]{Hao2020FL}
Hao, M.; Li, H.; Luo, X.; Xu, G.; Yang, H.; and Liu, S. 2020.
\newblock Efficient and Privacy-Enhanced Federated Learning for Industrial Artificial Intelligence.
\newblock \emph{IEEE Transactions on Industrial Informatics}, 16(10): 6532--6542.

\bibitem[{Hou et~al.(2020)Hou, Xie, Li, Wang, Lv, Xie, and Nan}]{Hou2020MultiStage}
Hou, X.; Xie, C.; Li, F.; Wang, J.; Lv, C.; Xie, G.; and Nan, Y. 2020.
\newblock A Triple-Stage Self-Guided Network for Kidney Tumor Segmentation.
\newblock In \emph{2020 IEEE 17th International Symposium on Biomedical Imaging (ISBI)}, 341--344.

\bibitem[{Hsiao et~al.(2022)Hsiao, Sun, Lin, Peng, Chen, Cheng, Yang, Yang, Wu, Lin, and Huang}]{OPLAB2022kidney}
Hsiao, C.-H.; Sun, T.-L.; Lin, P.-C.; Peng, T.-Y.; Chen, Y.-H.; Cheng, C.-Y.; Yang, F.-J.; Yang, S.-Y.; Wu, C.-H.; Lin, F. Y.-S.; and Huang, Y. 2022.
\newblock A deep learning-based precision volume calculation approach for kidney and tumor segmentation on computed tomography images.
\newblock \emph{Computer Methods and Programs in Biomedicine}, 221: 106861.

\bibitem[{Huang et~al.(2023)Huang, Bajaj, Li, Ye, Lin, Pugliese, Ramasamy, Gu, Wang, Torii, Dijkstra, Zhou, Bourantas, and Zhang}]{HUANG2023102922}
Huang, X.; Bajaj, R.; Li, Y.; Ye, X.; Lin, J.; Pugliese, F.; Ramasamy, A.; Gu, Y.; Wang, Y.; Torii, R.; Dijkstra, J.; Zhou, H.; Bourantas, C.~V.; and Zhang, Q. 2023.
\newblock POST-IVUS: A perceptual organisation-aware selective transformer framework for intravascular ultrasound segmentation.
\newblock \emph{Medical Image Analysis}, 89: 102922.

\bibitem[{Hui et~al.(2017)Hui, Cao, Zhang, Kole, Wang, Yu, Eakins, Sturek, Chen, and Cheng}]{Hui2017IVUS}
Hui, J.; Cao, Y.; Zhang, Y.; Kole, A.; Wang, P.; Yu, G.; Eakins, G.; Sturek, M.; Chen, W.; and Cheng, J.-X. 2017.
\newblock Real-Time intravascular photoacoustic-ultrasound imaging of lipid-laden plaque in human coronary artery at 16 frames per second.
\newblock \emph{Scientific Reports}, 7: 1417.

\bibitem[{Iatan et~al.(2023)Iatan, Guan, Humphries, Yeoh, and Mancini}]{10.1001/jamacardio.2023.2731}
Iatan, I.; Guan, M.; Humphries, K.~H.; Yeoh, E.; and Mancini, G. B.~J. 2023.
\newblock Atherosclerotic Coronary Plaque Regression and Risk of Adverse Cardiovascular Events: A Systematic Review and Updated Meta-Regression Analysis.
\newblock \emph{JAMA Cardiology}, 8(10): 937--945.

\bibitem[{Kumar et~al.(2021)Kumar, Khan, Kumar, Zakria, Golilarz, Zhang, Ting, Zheng, and Wang}]{Kumar2021FL}
Kumar, R.; Khan, A.~A.; Kumar, J.; Zakria; Golilarz, N.~A.; Zhang, S.; Ting, Y.; Zheng, C.; and Wang, W. 2021.
\newblock Blockchain-Federated-Learning and Deep Learning Models for {COVID}-19 Detection Using {CT} Imaging.
\newblock \emph{IEEE Sensors Journal}, 21(14): 16301--16314.

\bibitem[{Li et~al.(2022)Li, Zhang, Cao, Sun, Feng, Zhang, Yang, Li, and Liu}]{9844289}
Li, B.; Zhang, P.; Cao, Y.; Sun, L.; Feng, J.; Zhang, Y.; Yang, Q.; Li, Y.; and Liu, Z. 2022.
\newblock AIVUS: Guidewire Artifacts Inpainting for Intravascular Ultrasound Imaging With United Spatiotemporal Aggregation Learning.
\newblock \emph{IEEE Transactions on Computational Imaging}, 8: 679--692.

\bibitem[{Li et~al.(2021)Li, Shen, Chen, Chang, Lee, and Huang}]{Li2021IVUSSeg}
Li, Y.-C.; Shen, T.-Y.; Chen, C.-C.; Chang, W.-T.; Lee, P.-Y.; and Huang, C.-C.~J. 2021.
\newblock Automatic Detection of Atherosclerotic Plaque and Calcification From Intravascular Ultrasound Images by Using Deep Convolutional Neural Networks.
\newblock \emph{IEEE Transactions on Ultrasonics, Ferroelectrics, and Frequency Control}, 68(5): 1762--1772.

\bibitem[{Liu et~al.(2020)Liu, Yuan, Hu, Liang, Feng, Ai, and Zhang}]{Liu2020Pancreas}
Liu, S.; Yuan, X.; Hu, R.; Liang, S.; Feng, S.; Ai, Y.; and Zhang, Y. 2020.
\newblock Automatic Pancreas Segmentation via Coarse Location and Ensemble Learning.
\newblock \emph{IEEE Access}, 8: 2906--2914.

\bibitem[{Liu et~al.(2022)Liu, Chen, Zhao, Yu, Liu, Bao, Jiang, Nie, Xu, and Yang}]{Liu_Chen_Zhao_Yu_Liu_Bao_Jiang_Nie_Xu_Yang_2022}
Liu, Z.; Chen, Y.; Zhao, Y.; Yu, H.; Liu, Y.; Bao, R.; Jiang, J.; Nie, Z.; Xu, Q.; and Yang, Q. 2022.
\newblock Contribution-Aware Federated Learning for Smart Healthcare.
\newblock \emph{Proceedings of the AAAI Conference on Artificial Intelligence}, 36(11): 12396--12404.

\bibitem[{McMahan et~al.(2017)McMahan, Moore, Ramage, Hampson, and Arcas}]{McMahan2017FedAvg}
McMahan, B.; Moore, E.; Ramage, D.; Hampson, S.; and Arcas, B. A.~y. 2017.
\newblock {Communication-Efficient Learning of Deep Networks from Decentralized Data}.
\newblock In \emph{Proceedings of the 20th International Conference on Artificial Intelligence and Statistics}, volume~54, 1273--1282.

\bibitem[{Min et~al.(2022)Min, Chang, Andreini, Pontone, Guglielmo, Bax, Knaapen, Raman, Chazal, Freeman, Crabtree, and Earls}]{MIN2022PlaqueSeverity}
Min, J.~K.; Chang, H.-J.; Andreini, D.; Pontone, G.; Guglielmo, M.; Bax, J.~J.; Knaapen, P.; Raman, S.~V.; Chazal, R.~A.; Freeman, A.~M.; Crabtree, T.; and Earls, J.~P. 2022.
\newblock Coronary CTA plaque volume severity stages according to invasive coronary angiography and FFR.
\newblock \emph{Journal of Cardiovascular Computed Tomography}, 16(5): 415--422.

\bibitem[{Saito et~al.(2024)Saito, Kobayashi, Fujii, Sonoda, Tsujita, Hibi, Morino, Okura, Ikari, Kozuma, and Honye}]{Saito2024}
Saito, Y.; Kobayashi, Y.; Fujii, K.; Sonoda, S.; Tsujita, K.; Hibi, K.; Morino, Y.; Okura, H.; Ikari, Y.; Kozuma, K.; and Honye, J. 2024.
\newblock CVIT 2023 clinical expert consensus document on intravascular ultrasound.
\newblock \emph{Cardiovascular Intervention and Therapeutics}, 39(1): 1--14.

\bibitem[{Stone et~al.(2020)Stone, Maehara, Ali, Held, Matsumura, Kjøller-Hansen, Bøtker, Maeng, Engstrøm, Wiseth, Persson, Trovik, Jensen, James, Mintz, Dressler, Crowley, Ben-Yehuda, and Erlinge}]{STONE20202289}
Stone, G.~W.; Maehara, A.; Ali, Z.~A.; Held, C.; Matsumura, M.; Kjøller-Hansen, L.; Bøtker, H.~E.; Maeng, M.; Engstrøm, T.; Wiseth, R.; Persson, J.; Trovik, T.; Jensen, U.; James, S.~K.; Mintz, G.~S.; Dressler, O.; Crowley, A.; Ben-Yehuda, O.; and Erlinge, D. 2020.
\newblock Percutaneous Coronary Intervention for Vulnerable Coronary Atherosclerotic Plaque.
\newblock \emph{Journal of the American College of Cardiology}, 76(20): 2289--2301.

\bibitem[{Yan et~al.(2021)Yan, Wicaksana, Wang, Yang, and Cheng}]{Yan2021FL}
Yan, Z.; Wicaksana, J.; Wang, Z.; Yang, X.; and Cheng, K.-T. 2021.
\newblock Variation-Aware Federated Learning With Multi-Source Decentralized Medical Image Data.
\newblock \emph{IEEE Journal of Biomedical and Health Informatics}, 25(7): 2615--2628.

\bibitem[{Zhang et~al.(2021)Zhang, Zhou, Lu, Wang, Zhu, Sun, Wang, Lo, and Wang}]{Zhang2021FL}
Zhang, W.; Zhou, T.; Lu, Q.; Wang, X.; Zhu, C.; Sun, H.; Wang, Z.; Lo, S.~K.; and Wang, F.-Y. 2021.
\newblock Dynamic-Fusion-Based Federated Learning for {COVID}-19 Detection.
\newblock \emph{IEEE Internet of Things Journal}, 8(21): 15884--15891.

\bibitem[{Zhao et~al.(2021)Zhao, Fan, Yang, Wang, Li, and Yang}]{Zhao2021FL}
Zhao, B.; Fan, K.; Yang, K.; Wang, Z.; Li, H.; and Yang, Y. 2021.
\newblock Anonymous and Privacy-Preserving Federated Learning With Industrial Big Data.
\newblock \emph{IEEE Transactions on Industrial Informatics}, 17(9): 6314--6323.

\bibitem[{Zhou et~al.(2020)Zhou, Siddiquee, Tajbakhsh, and Liang}]{Zhou2020Unetplus}
Zhou, Z.; Siddiquee, M. M.~R.; Tajbakhsh, N.; and Liang, J. 2020.
\newblock UNet++: Redesigning Skip Connections to Exploit Multiscale Features in Image Segmentation.
\newblock \emph{IEEE Transactions on Medical Imaging}, 39(6): 1856--1867.

\bibitem[{Zhu et~al.(2022)Zhu, Gao, Zhao, Zhu, Nan, Tian, Dong, Jiang, Feng, Dai, and Zhou}]{doi:10.1177/01617346221114137}
Zhu, F.; Gao, Z.; Zhao, C.; Zhu, H.; Nan, J.; Tian, Y.; Dong, Y.; Jiang, J.; Feng, X.; Dai, N.; and Zhou, W. 2022.
\newblock A Deep Learning-based Method to Extract Lumen and Media-Adventitia in Intravascular Ultrasound Images.
\newblock \emph{Ultrasonic Imaging}, 44(5-6): 191--203.
\newblock PMID: 35861418.

\end{thebibliography}

\end{document}